 \documentclass[aps,prd,twocolumn,showpacs,amsmath,amssymb]{revtex4}
\usepackage{graphicx}
\usepackage{dcolumn}
\usepackage{bm}

\begin{document}

\newcommand{\greeksym}[1]{{\usefont{U}{psy}{m}{n}#1}}
\newcommand{\ual}{\mbox{\greeksym{a}}}
\newcommand{\unu}{\mbox{\greeksym{n}}}
\newcommand{\umu}{\mbox{\greeksym{m}}}
\newcommand{\utau}{\mbox{\greeksym{t}}}
\newcommand{\uth}{\mbox{\greeksym{q}}}
\newcommand{\uchi}{\mbox{\greeksym{c}}}
\newcommand{\urho}{\mbox{\greeksym{r}}}
\newcommand{\uepsilon}{\mbox{\greeksym{e}}}
\newcommand{\ugamma}{\mbox{\greeksym{g}}}
\newcommand{\udelta}{\mbox{\greeksym{d}}}
\newcommand{\uDelta}{\mbox{\greeksym{D}}}
\newcommand{\uOmega}{\mbox{\greeksym{W}}}
\newcommand{\usigma}{\mbox{\greeksym{s}}}
\newcommand{\uPi}{\mbox{\greeksym{P}}}

\title{Search for  dark matter WIMPs  using upward  
  through-going muons in   Super--Kamiokande}

\newcounter{foots}
\newcounter{notes}
\newcommand{\authoraticrr}{$^{a}$}
\newcommand{\authoratncen}{$^{b}$}
\newcommand{\authoratbu}{$^{c}$}
\newcommand{\authoratbnl}{$^{d}$}
\newcommand{\authoratuci}{$^{e}$}
\newcommand{\authoratcsu}{$^{f}$}
\newcommand{\authoratcnu}{$^{g}$}
\newcommand{\authoratgmu}{$^{h}$}
\newcommand{\authoratgifu}{$^{i}$}
\newcommand{\authoratuh}{$^{j}$}
\newcommand{\authoratui}{$^{k}$}
\newcommand{\authoratkek}{$^{l}$}
\newcommand{\authoratkobe}{$^{m}$}
\newcommand{\authoratkyoto}{$^{n}$}
\newcommand{\authoratlanl}{$^{o}$}
\newcommand{\authoratlsu}{$^{p}$}
\newcommand{\authoratumd}{$^{q}$}
\newcommand{\authoratmit}{$^{r}$}
\newcommand{\authoratduluth}{$^{s}$}
\newcommand{\authoratsuny}{$^{t}$}
\newcommand{\authoratnagoya}{$^{u}$}
\newcommand{\authoratniigata}{$^{v}$}
\newcommand{\authoratosaka}{$^{w}$}
\newcommand{\authoratseoul}{$^{x}$}
\newcommand{\authoratshizuokaseika}{$^{y}$}
\newcommand{\authoratshizuoka}{$^{z}$}
\newcommand{\authoratskku}{$^{aa}$}
\newcommand{\authorattohoku}{$^{bb}$}
\newcommand{\authorattokyo}{$^{cc}$}
\newcommand{\authorattokai}{$^{dd}$}
\newcommand{\authorattit}{$^{ee}$}
\newcommand{\authoratwarsaw}{$^{ff}$}
\newcommand{\authoratuw}{$^{gg}$}
\newcommand{\addressoficrr}[1]{$^{a}$ #1 }
\newcommand{\addressofncen}[1]{$^{b}$ #1 }
\newcommand{\addressofbu}[1]{$^{c}$ #1 }
\newcommand{\addressofbnl}[1]{$^{d}$ #1 }
\newcommand{\addressofuci}[1]{$^{e}$ #1 }
\newcommand{\addressofcnu}[1]{$^{g}$ #1 }
\newcommand{\addressofcsu}[1]{$^{f}$ #1 }
\newcommand{\addressofgmu}[1]{$^{h}$ #1 }
\newcommand{\addressofgifu}[1]{$^{i}$ #1 }
\newcommand{\addressofuh}[1]{$^{j}$ #1 }
\newcommand{\addressofui}[1]{$^{k}$ #1 }
\newcommand{\addressofkek}[1]{$^{l}$ #1 }
\newcommand{\addressofkobe}[1]{$^{m}$ #1 }
\newcommand{\addressofkyoto}[1]{$^{n}$ #1 }
\newcommand{\addressoflanl}[1]{$^{o}$ #1 }
\newcommand{\addressoflsu}[1]{$^{p}$ #1 }
\newcommand{\addressofumd}[1]{$^{q}$ #1 }
\newcommand{\addressofmit}[1]{$^{r}$ #1 }
\newcommand{\addressofduluth}[1]{$^{s}$ #1 }
\newcommand{\addressofsuny}[1]{$^{t}$ #1 }
\newcommand{\addressofnagoya}[1]{$^{u}$ #1 }
\newcommand{\addressofniigata}[1]{$^{v}$ #1 }
\newcommand{\addressofosaka}[1]{$^{w}$ #1 }
\newcommand{\addressofseoul}[1]{$^{x}$ #1 }
\newcommand{\addressofshizuokaseika}[1]{$^{y}$ #1 }
\newcommand{\addressofshizuoka}[1]{$^{z}$ #1 }
\newcommand{\addressofskku}[1]{$^{aa}$ #1 }
\newcommand{\addressoftohoku}[1]{$^{bb}$ #1 }
\newcommand{\addressoftokyo}[1]{$^{cc}$ #1 }
\newcommand{\addressoftokai}[1]{$^{dd}$ #1 }
\newcommand{\addressoftit}[1]{$^{ee}$ #1 }
\newcommand{\addressofwarsaw}[1]{$^{ff}$ #1 }
\newcommand{\addressofuw}[1]{$^{gg}$ #1 }
\author{
{\large The Super-Kamiokande Collaboration} \\
\bigskip
S.~Desai\authoratbu,
%
Y.~Ashie\authoraticrr,
S.~Fukuda\authoraticrr,
Y.~Fukuda\authoraticrr,
K.~Ishihara\authoraticrr,
Y.~Itow\authoraticrr,
Y.~Koshio\authoraticrr,
A.~Minamino\authoraticrr,
M.~Miura\authoraticrr,
S.~Moriyama\authoraticrr
% ...
\\
\smallskip
\smallskip
\footnotesize
\it
\addressoficrr{Kamioka Observatory, Institute for Cosmic Ray Research, University of Tokyo, Kamioka, Gifu, 506-1205, Japan}\\
\addressofncen{Research Center for Cosmic Neutrinos, Institute for Cosmic Ray Research, University of Tokyo, Kashiwa, Chiba 277-8582, Japan}\\
\addressofbu{Department of Physics, Boston University, Boston, MA 02215, USA}\\
\addressofbnl{Physics Department, Brookhaven National Laboratory, Upton, NY 11973, USA}\\
\addressofuci{Department of Physics and Astronomy, University of California, Irvine, Irvine, CA 92697-4575, USA }\\
\addressofcsu{Department of Physics, California State University, Dominguez Hills, Carson, CA 90747, USA}\\
\addressofcnu{Department of Physics, Chonnam National University, Kwangju 500-757, Korea}\\
\addressofgmu{Department of Physics, George Mason University, Fairfax, VA 22030, USA }\\
\addressofgifu{Department of Physics, Gifu University, Gifu, Gifu 501-1193, Japan}\\
\addressofuh{Department of Physics and Astronomy, University of Hawaii, Honolulu, HI 96822, USA}\\
\addressofui{Department of Physics, Indiana University, Bloomington,
  IN 47405-7105, USA} \\
\addressofkek{High Energy Accelerator Research Organization (KEK), Tsukuba, Ibaraki 305-0801, Japan }\\
\addressofkobe{Department of Physics, Kobe University, Kobe, Hyogo 657-8501, Japan}\\
\addressofkyoto{Department of Physics, Kyoto University, Kyoto 606-8502, Japan}\\
\addressoflanl{Physics Division, P-23, Los Alamos National Laboratory, Los Alamos, NM 87544, USA }\\
\addressoflsu{Department of Physics and Astronomy, Louisiana State University, Baton Rouge, LA 70803, USA }\\
\addressofumd{Department of Physics, University of Maryland, College Park, MD 20742, USA }\\
\addressofmit{Department of Physics, Massachusetts Institute of Technology, Cambridge, MA 02139, USA}\\
\addressofduluth{Department of Physics, University of Minnesota, Duluth, MN 55812-2496, USA}\\
\addressofsuny{Department of Physics and Astronomy, State University of New York, Stony Brook, NY 11794-3800, USA}\\
\addressofnagoya{Department of Physics, Nagoya University, Nagoya, Aichi 464-8602, Japan}\\
\addressofniigata{Department of Physics, Niigata University, Niigata, Niigata 950-2181, Japan }\\
\addressofosaka{Department of Physics, Osaka University, Toyonaka, Osaka 560-0043, Japan}\\
\addressofseoul{Department of Physics, Seoul National University, Seoul 151-742, Korea}\\
\addressofshizuokaseika{International and Cultural Studies, Shizuoka Seika College, Yaizu, Shizuoka, 425-8611, Japan}\\
\addressofshizuoka{Department of Systems Engineering, Shizuoka University, Hamamatsu, Shizuoka 432-8561, Japan}\\
\addressofskku{Department of Physics, Sungkyunkwan University, Suwon 440-746, Korea}\\
\addressoftohoku{Research Center for Neutrino Science, Tohoku University, Sendai, Miyagi 980-8578, Japan}\\
\addressoftokyo{The University of Tokyo, Tokyo 113-0033, Japan }\\
\addressoftokai{Department of Physics, Tokai University, Hiratsuka, Kanagawa 259-1292, Japan}\\
\addressoftit{Department of Physics, Tokyo Institute for Technology, Meguro, Tokyo 152-8551, Japan }\\
\addressofwarsaw{Institute of Experimental Physics, Warsaw University, 00-681 Warsaw, Poland }\\
\addressofuw{Department of Physics, University of Washington, Seattle, WA 98195-1560, USA}\\
}
\affiliation{}

\begin{abstract}
   We present the results of indirect searches for Weakly Interacting
  Massive Particles (WIMPs), with 1679.6 live days of data from the
 Super-Kamiokande detector using   neutrino-induced upward through-going muons. The search is performed
  by looking for an excess  of high energy muon
  neutrinos from WIMP annihilations in  the Sun,
  the core of the Earth, and the Galactic Center, as compared to the
  number expected from the atmospheric neutrino background. No
  statistically significant excess was seen. We calculate the flux limits in various angular
  cones around each of the above celestial objects.  We obtain
 conservative  model-independent upper limits on the WIMP-nucleon cross-section as a function
 of WIMP mass, and compare these results with the corresponding results from
  direct dark matter detection experiments.

\end{abstract}
\maketitle

\section{Introduction}
There is growing  evidence which indicates that the
universe contains, at all length-scales,  ``dark matter''  which does not emit or absorb
electromagnetic radiation at any known wavelength, but manifests itself only
through gravity~\cite{Trimble87}. Recent data from WMAP and other
CMB experiments, large-scale structure surveys, Lyman-$\alpha$
forest,  Type 1a supernovae  etc.~\cite{WMAP,Tegmark,Perlmutter,Riess,Tonry} 
point with increasing accuracy towards  a standard cosmological 
model~\cite{Turner02,Silk02}, in which the universe is flat, with about 
5\% baryons, 25\% non-baryonic dark matter,   and about 70\% dark energy. 
In this model,  the non-baryonic dark matter consists of 
``Cold Dark Matter'' particles, which decoupled from
rest of the matter and radiation while moving at non-relativistic 
velocities~\cite{Peebles,Primack}.


Weakly Interacting Massive Particles (WIMPs) are one of the
 plausible  cold dark matter candidates~\cite{Jungman96,Bergstrom00}.  WIMPs are
 stable particles which are predicted to occur  in extensions of the
 standard model, such as  supersymmetry.  WIMPs undergo only weak-scale interactions with matter,
and could have masses in the GeV-TeV range.  If WIMPs exist,
their relic abundance (which is governed by electroweak scale
interactions) is remarkably close to the inferred density of dark matter
in the universe~\cite{Weinberg77}.  The lightest supersymmetric particle (LSP)
of supersymmetric theories is the most theoretically
well developed WIMP candidate~\cite{Goldberg83,Ellis84}.  Most SUSY
theories contain a multiplicatively conserved quantum number called
R-parity, where R-parity is equal to~1 for ordinary standard model particles,
and is equal to~$-1$ for their supersymmetric partners~\cite{Jungman96}.
If R-parity is conserved, superpartners can form and decay only in
pairs. Thus, the LSP is stable and hence should be present in the Universe
as a cosmological relic from the Big Bang.  The most likely candidate
for this LSP is the neutralino~\cite{Ellis84}.

The neutralino $\tilde{\chi}$ is a linear combination 
of the supersymmetric particles 
that mix after electroweak-symmetry breaking:
\begin{equation}
\tilde{\chi}= a_{1}\tilde{\gamma} + a_{2}\tilde{Z} +
a_{3}\tilde{H}_{1} + a_{4}\tilde{H}_{2},  
\end{equation}
\noindent where $\tilde{\gamma}$ and $\tilde{Z}$ are the 
supersymmetric partners of the photon and the $Z$ boson, and
$\tilde{H}_{1}$ and $\tilde{H}_{2}$ are the supersymmetric partners of
the Higgs bosons (Higgsino).  
The current lower limit on neutralino mass, without assuming anything about
the SUSY breaking mechanism, and taking into account the results from 
LEP2 data,  is about 18~GeV~\cite{Hooper02}. The upper limit on neutralino mass
depends on the model assumed for SUSY breaking,
and ranges from   500 GeV in some models~\cite{Ellis03} and up to 10~TeV in other models~\cite{Gondolo97}. 

In this paper, we perform an indirect search for WIMP dark matter  using 
the Super--Kamiokande detector, by looking for a high energy neutrino signal resulting from
WIMP annihilation in the Earth, the Sun, and the Galactic Center.  The
signature would be an excess of neutrino-induced events coming from the
direction of these objects over the background expected from atmospheric
neutrinos.  Such a search is complementary  to direct detection 
experiments, which
look for direct interaction of WIMPs with a nucleus in a
low background detector.  However, both direct and indirect detection
experiments probe the coupling of WIMPs to nuclei.  We then compare our
results with those of direct detection of dark matter experiments.

\section{Indirect WIMP Searches using Neutrino-Induced Muons}

If WIMPs constitute the dark matter in our
galactic halo, they will accumulate in the Sun~\cite{Press} and the Earth~\cite{Freese,Krauss}.  When their
orbits pass though a celestial body, the WIMPs have a small but finite
probability of elastically scattering with a nucleus of that body. If
their final velocity after scattering is less than the escape velocity,
they become gravitationally trapped and eventually settle into the center
of that body.  WIMPs which have accumulated this way  annihilate in pairs,
primarily into $\tau$ leptons, $b$, $c$ and $t$ quarks, gauge bosons, and
Higgs bosons, depending upon their mass and composition.  As the WIMP
density increases in the core, the annihilation rate increases until
equilibrium is achieved between capture and annihilation, making the
annihilation (each of which involves two captured WIMPs) rate half of
the capture rate.  High energy muon neutrinos
are produced by the decay of the annihilation products.  The expected
neutrino fluxes from the capture and annihilation of WIMPs in the Sun
and the Earth depend upon several astrophysical parameters: the WIMP
mean halo velocity ($v_{\chi} \sim 220$~km~s$^{-1}$), the WIMP local density
($\rho_{\chi} \sim 0.3$~GeV~cm$^{-3}$), the WIMP-nucleon scattering
cross-section; and the mass and escape velocity of the celestial body.
WIMPs undergo two kinds of interactions with the celestial bodies: axial
vector interactions in which WIMPs couple to the spin of the nucleus;
and scalar interactions in which WIMPs couple to the mass of the
nucleus~\cite{Witten}.    For the Earth, WIMPs mainly undergo  scalar
interactions, since the abundance of nuclei with odd mass number (which
are needed for spin coupling) is
extremely small~\cite{Anderson}. In the Sun,  WIMPs interact through
both scalar
as well as  axial vector interactions (due to the large abundance of hydrogen). There are many calculations of
the expected neutrino fluxes from WIMP capture and annihilation in the
Sun and the Earth~\cite{Freese,Krauss,Silk,Steigman,RS88,Jungman95,Bottino95,Berezinsky,Berg,Berg98,Feng,Halzen}.  

It has been also pointed out that  cold dark matter 
near the Galactic Center can be accreted by the central black
hole resulting in a dense spike in the density distribution~\cite{Gondolo}.
WIMP annihilations in this spike would make it a compact source of high energy
neutrinos. The flux of these neutrinos depends on the density profile of
the inner dark matter halo.  Halos with isothermal or finite cores
produce a negligible flux of neutrinos, while halos with inner density
cusps produce an appreciable flux of neutrinos~\cite{Gondolo}.

Energetic neutrinos coming from WIMP annihilation in the Sun, the
Earth, and the Galactic Center could be detected in neutrino detectors.
The mean neutrino energy ranges from 1/3 to 1/2 the mass of the
WIMP, i.e. from about 5 GeV to 5 TeV. 
In this energy range, neutrino-induced upward through-going  muons from
charged current  interactions provide the most effective signatures in
Super-Kamiokande.
\section{WIMP Analysis with Super--K}

The Super--Kamiokande (``Super--K'') detector is a 50,000~ton water
Cherenkov detector, located in the Kamioka-Mozumi mine in Japan with
1000~m rock overburden.
For this analysis, Super-K  consisted of an inner detector with 11,146
inward-facing 50~cm  photomultiplier tubes (PMTs) and an outer
detector equipped with 1885 outward-facing 20~cm PMTs, serving
as a cosmic ray veto counter.  More details about the detector can be
found in Ref.~\cite{SKnim}.  The data used in this analysis was taken
from April 1996 to July 2001, corresponding to 1679.6 days of detector
livetime.

Upward through-going muons in Super--K are mainly produced by interactions of
atmospheric $\nu_{\mu}$ in the rock around the detector and are
energetic enough to cross the entire detector.  The effective target
volume extends outward for many tens of meters into the surrounding rock
and increases with the energy of the incoming neutrino, as the high
energy muons resulting from these interactions can travel longer
distances to reach the detector.  Thus, upward through-going muons
represent the highest energy portion of the atmospheric neutrino
spectrum observed by Super--K, and the corresponding parent neutrino energy
spectrum peaks at $\sim $ 100 GeV~\cite{SK99}.  The downward-going cosmic ray
muon rate in Super--K is 3~Hz, making it impossible to distinguish any
downward-going neutrino-induced muons from this large background of
downward-going cosmic ray muons. Hence we restrict our analysis to upward-going
muons.

Reconstruction of a muon  event is performed using the charge and timing
information recorded by each hit PMT. 
 Muons are required to have $\geq 7$~meters measured
path length (E$_{\mu} > 1.6$~GeV) in the inner detector. The
effective detector  area for upward through-going muons with this
 path length cut is $\sim 1200~\mbox{m}^2$.  For each event, we obtain the arrival
direction and time. After a 
final precision fit and visual scan, 1892 upward through-going muon events have been
observed. The detector angular resolution for upward through-going muons is about $1^{\circ}$.
More details of the data reduction procedure for selecting upward 
going muons can be found in Ref.~\cite{SKupmu}.  
The sample described above is contaminated by some
downward-going cosmic ray muons close to the horizon, which appear to be
upward-going due to the tracking angular resolution of the detector and
multiple Coulomb scattering in the rock.  The total number of such
non-$\nu$ background events has been estimated to be $ 14.4 \pm 9.4$,
all contained in the $-0.1 < \cos{\Theta} < 0$ bin, where ${\Theta}$ is the
zenith angle.  With the 7~m path-length cut, the contamination from photoproduced upward going pions
from nearby downward-going cosmic ray muons is estimated to be $< 1 \%
$~\cite{MACRO98}.

The expected background  due to interactions of
atmospheric $\nu$'s in the rock below the detector is evaluated with
Monte Carlo simulations. These simulations, which were generated
using the {\tt Nuance } neutrino simulation package~\cite{nuance}, use the Bartol atmospheric
${\nu}$ flux~\cite{Agarwal}, the GRV-94 parton distribution
function~\cite{GRV-94}, energy loss mechanisms of muons in rock
from Ref.~\cite{Lipari93}, and a GEANT-based detector simulation.
There is a 20$\%$ uncertainty in the prediction of absolute upward-going
muon flux.
Analysis of the most recent Super-K data~\cite{SK98,SK03} of upward going
muons and contained events is consistent with ${\nu}_{\mu} \rightarrow
{\nu}_{\tau}$ oscillations with best fit values given by : $\sin^2 2{\theta} \simeq 1$ and 
$\Delta
\mbox{m}^2 \simeq 2 \times 10^{-3} \mbox{eV}^2$.  For
evaluating our background, we suppress the atmospheric muon neutrino flux
due to oscillations from ${\nu}_{\mu}$ to ${\nu}_{\tau}$ using these
oscillation parameters.  

\begin{figure}

\center{\includegraphics[width=7.5cm]{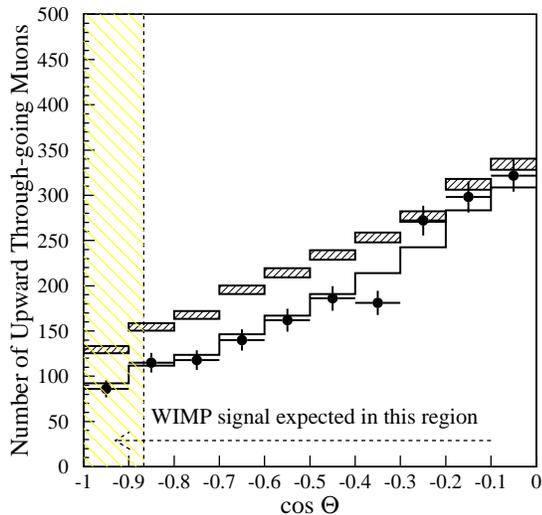}}
\caption{ \small Zenith angle distribution of upward through-going
muons with respect to the center of the Earth along with comparison
against the expected flux. The black circles indicate observed data
along with statistical uncertainties.
Hatched regions indicates the background from atmospheric neutrinos,
the solid lines indicate the atmospheric neutrino background after
taking into account neutrino oscillations with:  $\sin^2 2{\theta} = 1.0$ and
    $\Delta \mbox{m}^2 =2 \times 10^{-3} \mbox{eV}^{2}. $ 
 The hatched region in $-1.0<cos(\Theta)<-0.866$ range indicates 
the angular region where WIMP searches were done.}
\label{fig-1} 
\end{figure}

\begin{figure}[hbt]
\center{\includegraphics[width=7.5cm]{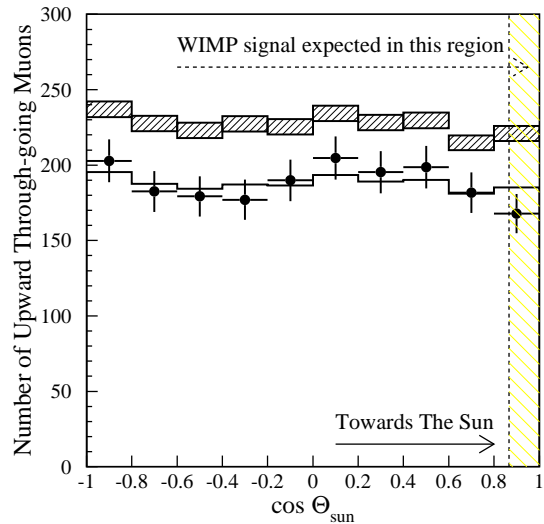}}
\caption{\small Angular distribution of upward through-going muons with
  respect to the Sun. All symbols are same as in Fig.~\ref{fig-1}.}

\label{fig-2} 
\end{figure}

\begin{figure}[hbt]
\includegraphics[width=7.5cm]{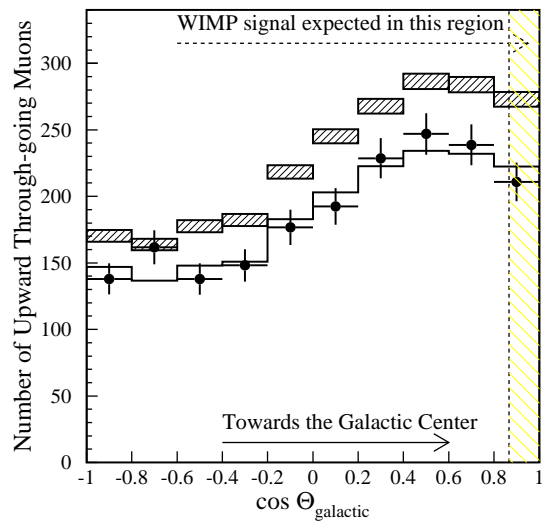}
\caption{\small Angular distribution of upward through-going muons with 
respect to the Galactic Center.  
All symbols are same as in Fig.~\ref{fig-1}.}

\label{fig-3} 
\end{figure}

The distribution of upward through-going muons
with respect to the Earth is shown in Fig.~\ref{fig-1}.  For the Sun,
there is  an additional  background of high energy neutrinos resulting from
cosmic ray interactions in the Sun, but this is 
negligible compared to the observed atmospheric ${\nu}$ flux and hence can be
neglected~\cite{Gaisser91,Ingel}. Normalization was done by
constraining the total number of Monte Carlo events to be equal to the
observed events, after taking oscillations into account.
In order to compare the
expected and observed distribution of upward through-going muon events
with respect to the Sun and Galactic Center, each Monte Carlo event was
assigned a time sampled from the observed upward
through-going muon arrival time, in order to match the livetime distribution
of the observed events.
This procedure allows us to obtain the right ascension and declination
for every Monte Carlo event,  from which the angle
between the upward muon and any celestial object  can be estimated.
  The distribution of upward muons with respect to the Sun and the
Galactic Center is shown in Figs.~\ref{fig-2} and~\ref{fig-3}
respectively.  

\begin{figure}
\center{\includegraphics[width=7.5cm]{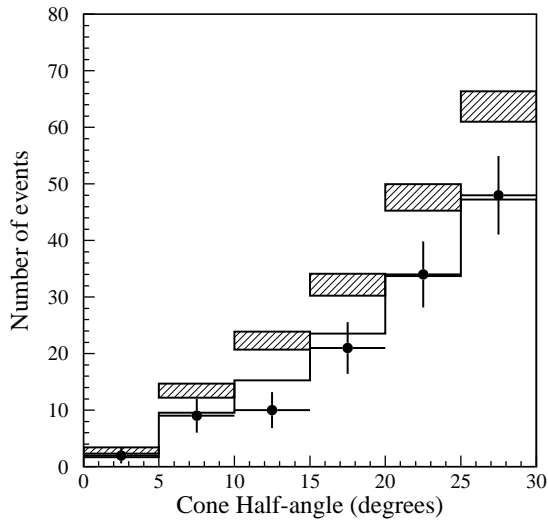}}
\caption{\small  The expanded view  of the zenith angle distribution of
 upward through-going muons around the Earth's center (Fig.~\ref{fig-1}). 
All symbols are same as in Fig.~\ref{fig-1}.}
\label{fig1-zoom} 
\end{figure}

\begin{figure}
\center{\includegraphics[width=7.5cm]{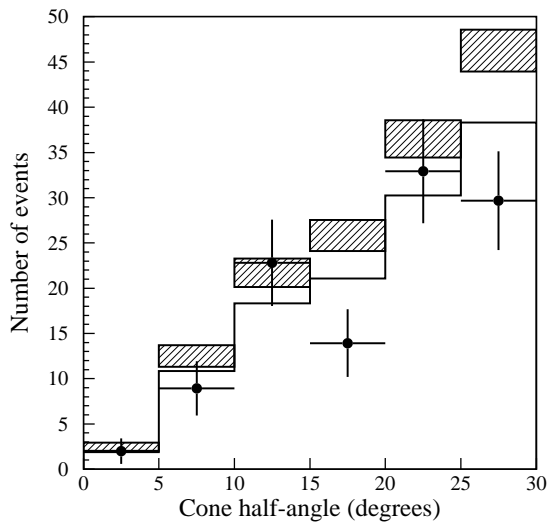}}
\caption{\small  The  expanded view  of the angular distribution of  
upward through-going muons around the 
Sun (Fig.~\ref{fig-2}). All symbols are same as in Fig.~\ref{fig-1}.}
\label{fig2-zoom} 
\end{figure}

\begin{figure}
\center{\includegraphics[width=7.5cm]{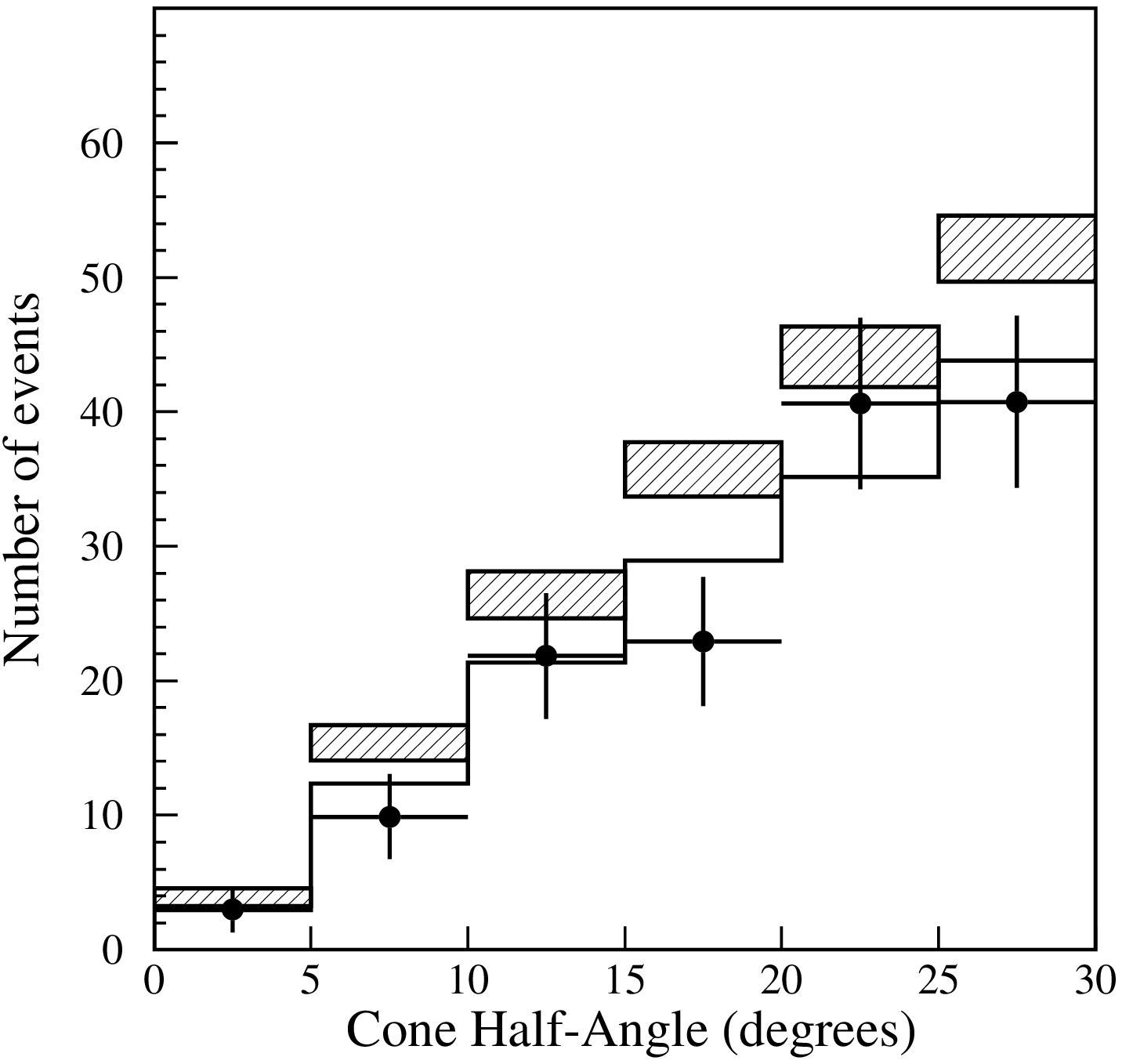}}
\caption{\small The  expanded view  of the angular distribution of  
upward through-going muons  around the Galactic Center (Fig.~\ref{fig-3}).
All symbols are  same as in Fig.~\ref{fig-1}.}
\label{fig3-zoom} 
\end{figure}

\section {RESULTS  FROM WIMP SEARCHES }
We searched for a statistically significant excess of muons in cones
about the potential source of neutrinos, with half angles ranging from 5
to 30 degrees.  This ensures that we catch $90\%$ of the signal
for a wide range of WIMP masses, as smaller masses produce a wider
angular distribution of neutrinos. 
Searching in different cone angles allows us to optimize the signal to background 
ratio as a function of  neutralino mass.

The distribution of
data and Monte Carlo (both with and without oscillations) in different
angular regions ranging from 5 to 30 degrees around the center  of
the Earth, the Sun and the Galactic Center is shown in
 Figs.~\ref{fig1-zoom},~\ref{fig2-zoom} and ~\ref{fig3-zoom} respectively. 
There was no statistically significant excess that was seen 
in any half angle
cones up to $30^{\circ}$.  We calculate the flux limit of excess neutrino-induced muons in
each of the cones.
The flux limit is given by :
\begin{equation}
\Phi(90\% \, CL) = \frac{N_{90}}{E}
\end{equation}
\noindent where $N_{90}$ is the upper Poissonian limit (90$\%$
CL) given the number of measured events and expected background
\cite{PDB} (due to atmospheric neutrinos after oscillations), and 
$ E$ is the exposure given by equation:
\begin{equation}
\label{EXP}
 E = \epsilon \times A \times T 
\end{equation}
\noindent where $A$  is the detector area in the direction
of the expected signal ; $\epsilon$ is the
detector efficiency which is $\approx 100\%$ for upward through-going
muons; and $T$ is the experimental livetime.

\begin{figure}[hbt]
\center{\includegraphics[width=7.5cm]{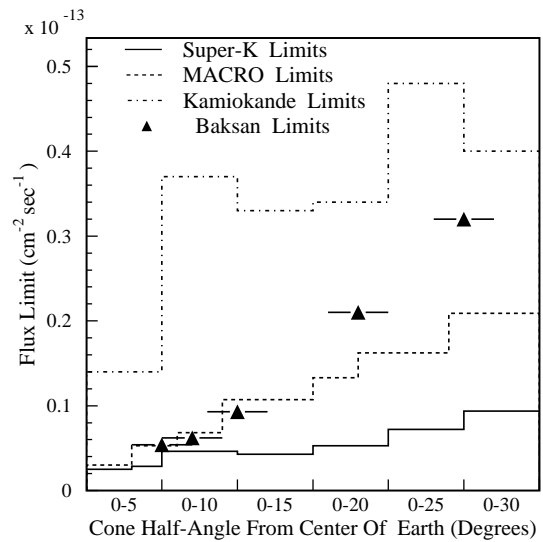}}
\caption{\small Comparison of Super-K 90 \% CL excess neutrino-induced
  upward muon flux limits from the Earth in cone half-angles ranging
from $5^{\circ}$ to $30^{\circ}$ along  with those from   other experiments.}
\label{fig-4} 
\end{figure}

\begin{figure}[hbt]
\center{\includegraphics[width=7.5cm]{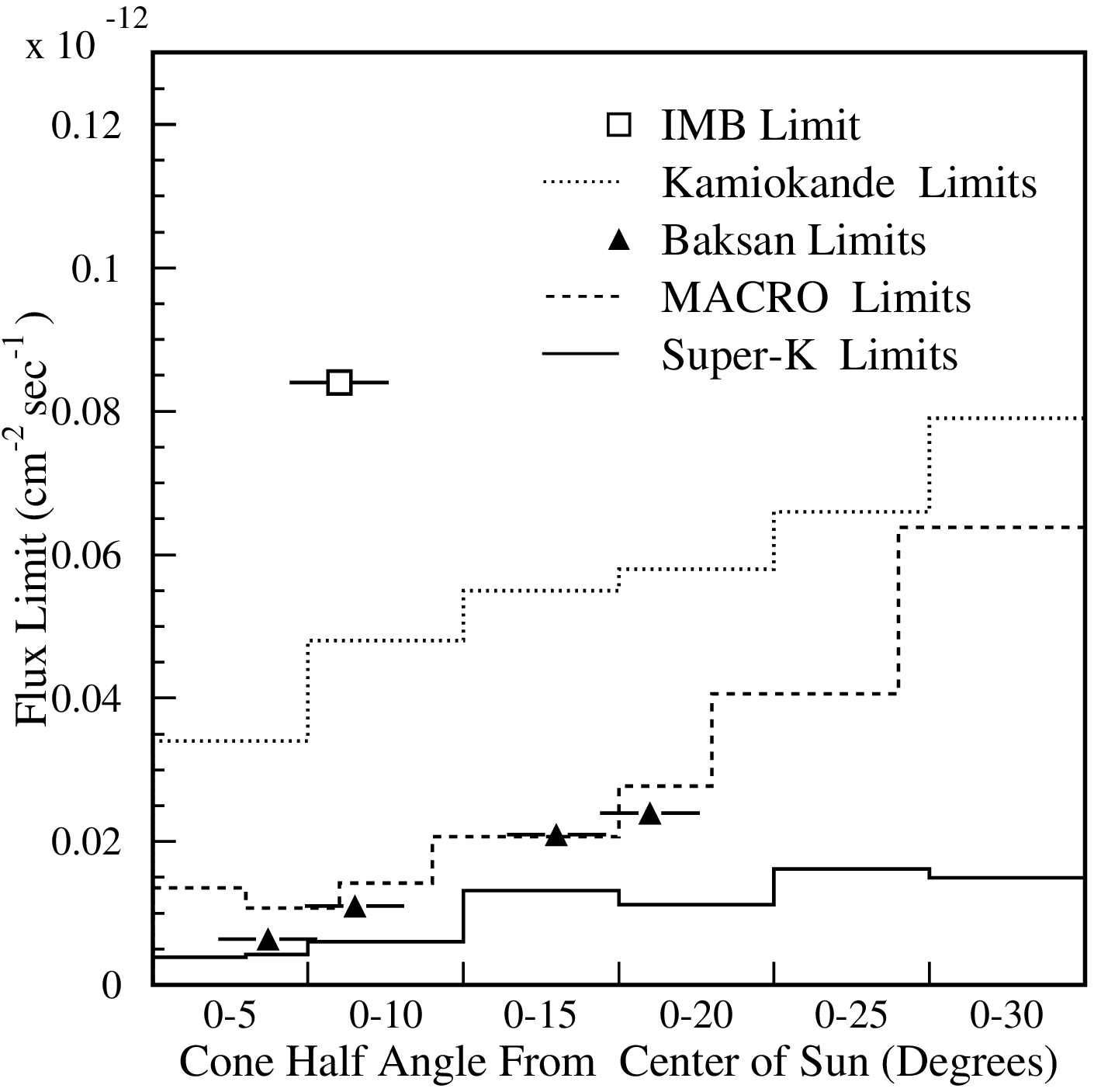}}
\caption{\small Comparison of Super-K 90 \% CL excess neutrino-induced
  upward muon flux limits from  the Sun   in cone half-angles ranging
from $5^{\circ}$ to $30^{\circ}$  along  with those from  other experiments.}
\label{fig-5} 
\end{figure}

\begin{figure}[hbt]
\center{\includegraphics[width=7.5cm]{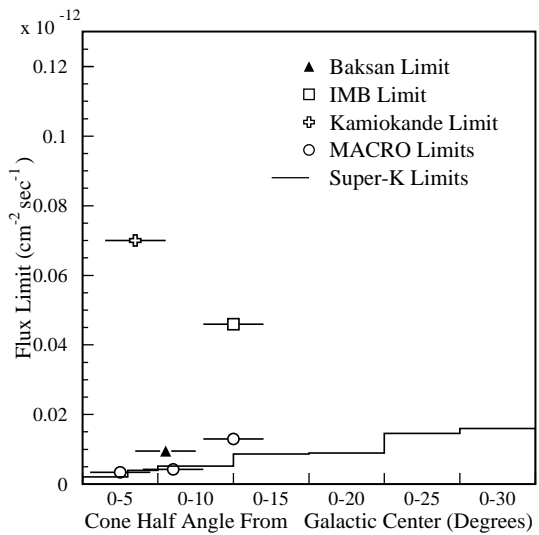}}
\caption{\small Comparison of Super-K 90 \% CL excess neutrino-induced
  upward muon flux limits from the Galactic Center in  cone
  half-angles ranging from $5^{\circ}$ to $30^{\circ}$ along with those from other experiments.}
\label{fig-6} 
\end{figure}

The comparison of Super-K flux limits with  searches 
by other experiments is shown in  Figs.~\ref{fig-4}--\ref{fig-6}. All the 
other experiments  have muon energy 
thresholds around 1 GeV. The upward muon flux limits for the Earth and the Sun by
 MACRO, Kamiokande, Baksan,  and IMB are
in Refs.~\cite{MACRO99,Mori91,Baksan99,IMB86} 
respectively, and  those  for the Galactic Center by the
above  detectors are in Refs.~\cite{MACRO2,Kamiokande89,Baksan2,IMB87}  
respectively. We also find that varying the oscillation parameters
and choosing values corresponding to the boundaries of the 90 \% 
confidence level
allowed region does not change the flux limits within the first 20 degrees 
from the celestial object. 
Only in the largest half angle cones (half angle 30 degrees) does the
flux limit vary by as much as $\pm$10 \%.
In this paper all results
are reported with the above best-fit oscillation parameters.


   Once WIMPs are captured in the Sun and the Earth they settle to the
core  with an isothermal distribution equal to the core
temperature of the Sun or of the Earth~\cite{Gould87}. While the Sun is  
effectively a point source of energetic 
neutrinos resulting from WIMP annihilations, the Earth is not a point
source  for WIMPs 
with mass less than  50 GeV.  For the Earth, the angular shape of
the annihilation region has been estimated to be :~\cite{Bottino95,Gould87}
\nopagebreak

\begin{equation}
G(\Theta) \simeq 4 m_{\chi} \beta e^{-2 m_{\chi} \beta \sin^{2}\Theta}
\end{equation}
\nopagebreak
\noindent where $\Theta$ is the nadir angle, $\beta$ is a
parameter depending on the central temperature,  the
central density  and radius of the Earth. In addition, muons deviate from the 
incoming direction of their parent
neutrino due to multiple coulomb scattering of the muon.

We used the Monte Carlo simulations done in Ref.~\cite{Mori91} to calculate 
the angular windows for the Sun and the Earth, which contain 90$\%$ of 
the signal for various neutralino masses, taking into account the spread
because of neutrino physics as well as the finite angular size of the WIMP
annihilation region for the Earth.  The 
simulations assumed that 80$\%$  of the annihilation products are
from $b\bar{b}$, 10$\%$ from $ c\bar{c}$ and 10$\%$ from $\tau^{+} \tau^{-}$.
However the neutrino-muon scattering angle is mainly dependent on the 
neutralino mass, and does not change much with different branching
ratios~\cite{MACRO99}. For the case of the  Earth, 
cones of half-angle $2^{\circ}$ and $22^{\circ}$ contain 90\% of the
signal from WIMPs with masses of 10~TeV and 18~GeV respectively. 
The corresponding numbers for the Sun are  $1.5^{\circ}$
and $19^{\circ}$ for WIMPs with masses of 10~TeV and 18~GeV respectively.
The corresponding cone half-angle for other intermediate masses can be found
 in Ref.~\cite{Mori91}.

\begin{figure}[hbt]
\center{\includegraphics[width=7.5cm]{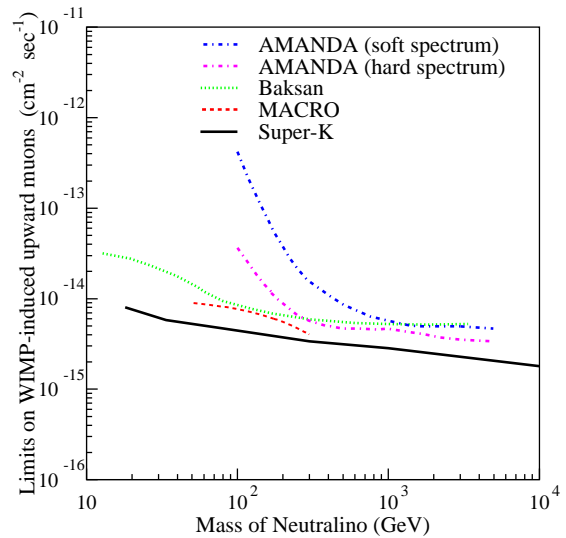}}
\caption{ \small Super-K WIMP-induced  upward muon flux limits from 
Earth  as a function  of  WIMP mass along with those from other 
detectors. }
\label{fig-7} 
\end{figure}

\begin{figure}[hbt]
\center{\includegraphics[width=7.5cm]{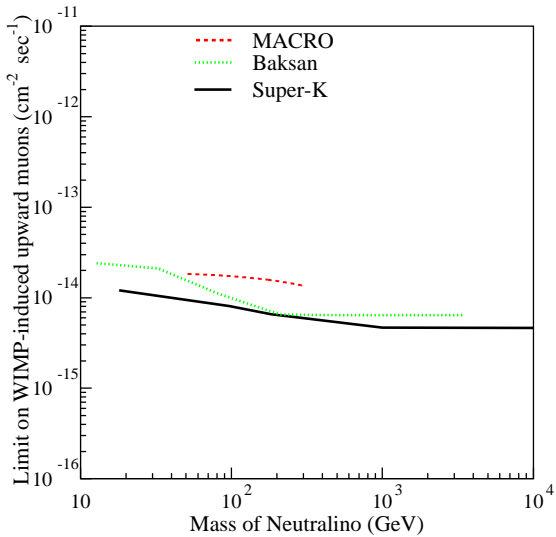}}
\caption{\small  Super-K WIMP-induced  upward muon flux 
limits from  Sun  as a function  of  WIMP mass along with 
comparison with other detectors.}
\label{fig-7a} 
\end{figure}

Using these windows, 90$\%$ confidence level flux limits can
be calculated  as a function of neutralino mass, using cones which collect
90$\%$ of expected signal for any given mass and after correcting for the 
collection efficiency of the cones. These flux limits as a 
function of neutralino mass are shown in Figs.~\ref{fig-7} and~\ref{fig-7a} for the Earth and the Sun respectively,  along with similar 
limits from AMANDA~\cite{Amanda}, BAKSAN~\cite{Baksan2},
MACRO~\cite{MACRO99}. The Super-K limits have been plotted from 18 GeV
up to 10 TeV. The lower limit of 18 GeV is the minimum WIMP
mass  for which at least 90\% of the upgoing muon signal would be
through-going, rather than stopping upward muons.
These limits can be compared with theoretical expectations from SUSY models.


\begin{figure}[hbt]
\center{\includegraphics[width=7.5cm]{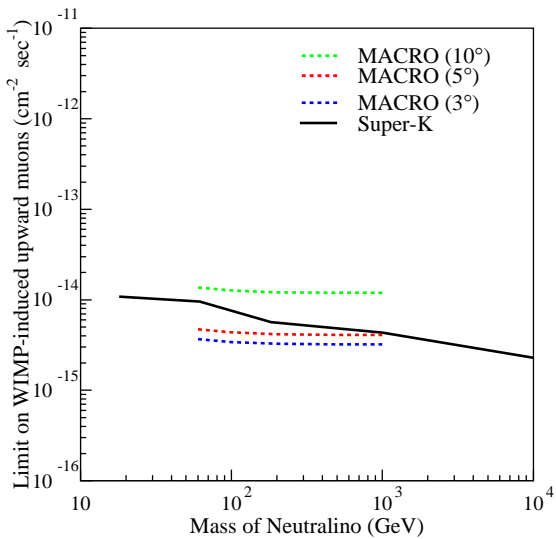}}
\caption{\small Super-K WIMP-induced  upward muon flux limits from
Galactic Center  as a function  of  neutralino mass along with
corresponding limit from MACRO.}
\label{fig-8} 
\end{figure}

For the case of the galactic center,  the apparent  size of the annihilation 
region is less than 0.05$^{\circ}$~\cite{Gondolo}. Hence the Galactic Center 
can be considered a point source for WIMP annihilation, and the angular
point spread function  is same as that for the Sun.
  These flux limits, as a 
function of neutralino mass for the Galactic Center (which also have been
plotted from 18~GeV up to 10~TeV),  are shown in
Fig.~\ref{fig-8}.

\section{Comparison with direct detection experiments}
Direct detection  experiments seek to observe the keV  energy recoil
in a low-background  detector, when a WIMP elastically scatters 
from a nucleus therein. On the other hand, indirect detection experiments
involve a search for energetic neutrinos produced by annihilation of WIMPs,
which have been trapped in the center of the Sun and the Earth. Rates for
both 
techniques depend primarily upon the WIMP-nucleon 
cross-section and mass.  
It is possible to compare direct and indirect
experiments by making realistic assumptions about some of 
the model-dependencies~\cite{marc96}.  
We sketch some of the details and outline how we obtain limits on 
WIMP-nucleon cross-section using these results.

When one
calculates the ratio of direct detection rates to the flux of
upward-going muons from WIMP annihilation, the main model-dependent term
which cannot be canceled and depends on details of the SUSY spectrum 
is in the quantity defined in Ref.~\cite{marc96} as $\xi(m_\chi)$; this is
the second moment of the neutrino energy spectrum from a given
annihilation channel, scaled by the branching ratio of WIMP
to that annihilation channel: 
\begin{equation}      
\begin{array}{c}
\xi(m{_\chi}) = \sum_{F} B_F [3.47 <Nz^2>_{F,\nu}(m_{\chi}) \\ + 2.08
<Nz^2>_{F,\overline{\nu}}(m_{\chi})].
\end{array}
\label{xi}
\end{equation}
Here, the sum is over all annihilation channels, F, available to any specific
neutralino candidate,  $B_{F}$ is the branching ratio for
annihilation, and  $<Nz^2> $ is the scaled second moment of the neutrino energy spectrum
from final state F for a given neutralino mass. All other terms  are
functions of only the neutralino mass and the elastic scattering 
cross-section.
As argued in Ref.~\cite{Jungman96}, an upper and lower bound can be
estimated for $\xi(m_{\chi})$  depending on the neutralino mass. 
For a WIMP heavier than the top quark, the lower limit
comes from annihilation to gauge bosons. For WIMPs lighter than 
the $W$ boson, the lower limit comes from annihilation to $b\bar{b}$,
and for intermediate mass WIMPs the lower limit comes from annihilation 
to $\tau\bar{\tau}$ pairs. 
Using these assumptions, maximum
values of the ratio of direct to indirect fluxes have been calculated 
for WIMPs with pure  scalar couplings, and for WIMPs with pure
 axial vector coupling. These ratios have been  plotted in Figure (2) of Ref~\cite{marc96}. In comparing
direct and indirect searches for scalar-coupled WIMPs, the sum of WIMP-induced
upward muon flux from  the Sun and the Earth has been taken into account, whereas for
  WIMPs with axial vector couplings only the   WIMP-induced
upward muon flux  from the Sun has been considered.
To get a rough idea of the sensitivities of the direct and indirect searches, 
 the event rate in a 1~kg  Germanium detector is 
equivalent to the event rate in $10^{4}-10^{6}$ \mbox{$m^2$} of upward muon 
detector for a WIMP with scalar
couplings, and the event rate in a 50-gm hydrogen detector is 
roughly same as that in a $10-500$ \mbox{$m^2$} muon detector. Using the maximum
inferred ratios  of direct to 
indirect detection rates,  we can compare our results with those  from direct
detection experiments  as we shall describe below. Before that we shall 
briefly describe results from some of the direct dark matter detection 
experiments.

The  DAMA experiment at Gran Sasso has claimed  that their data,
collected over 7 annual cycles corresponding to 107731 kg.day
exposure, contain an annual modulation which is consistent with
the possible presence of scalar-coupled WIMPs~\cite{DAMA00,DAMA2003},
with the best fit values being : $M_w$ = 52 GeV 
and $\sigma_p= 7.2 \times 10^{-6}$ pb using standard
astrophysical assumptions~\cite{DAMA00}.  
The CDMS~\cite{CDMS,CDMS02,CDMS03,CDMS04}  and EDELWEISS~\cite{Edelweiss}  
experiments, which employ Ge and Si detectors, and the 
ZEPLIN~\cite{ZEPLIN} experiment, which uses liquid Xe, do not see 
any WIMP signal. They rule out most or all of  the DAMA allowed region
at more than $99\%$ CL.

 Using the maximum inferred ratio  of the direct-to-indirect detection
 rates, we get 
 conservative model independent limits on 
neutralino-nucleon  spin-independent as well as spin-dependent 
cross-sections (on protons) from the Super-K
flux limits, and compare them with the results of direct-detection
 experiments, since we have obtained $90\%$  CL  upper flux limit.

\begin{figure}[hbt]
\center{\includegraphics[width=7.5cm]{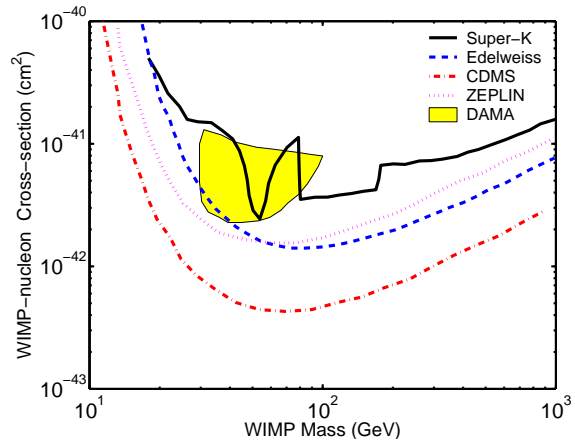}}
\caption{Super-K 90 \% CL exclusion region in WIMP parameter space 
(solid   line)  for a WIMP with scalar coupling obtained using limits
on WIMP-induced muons from the Sun and the Earth. Also shown are
the DAMA  3$\sigma$ allowed region 
(filled), 90 \% CL exclusion region from CDMS (dot-dashed), 
EDELWEISS (dashed), and ZEPLIN (dotted).}
\label{figsksi} 
\end{figure}

To get limits on the WIMP-nucleon cross-section for
a WIMP with scalar coupling,   we first calculate 
the combined WIMP flux limits from the Sun and the Earth as a  
function of WIMP mass and solve the following equation
to calculate the upper limit on WIMP-nucleon cross-section:
\begin{equation}
\mbox{Max Ratio (M)} = \frac{\mbox{Direct Detection Rate ( M, $\sigma$
    )}}{\mbox{Super-K limit (M)}}
\label{ddrate}
\end{equation}
where the quantity on the left   is the maximum calculated ratio of direct to indirect 
detection rates for a WIMP with scalar coupling, and the numerator on the
right hand side indicates the event rate in a direct detection experiment as
a function of WIMP mass and cross-section. The  denominator on the right hand side
is the combined Super-K limit from the  Sun and the Earth as a
function of WIMP mass. To obtain the combined flux limit, we first
calculate the total number of observed events for each WIMP
mass.  This is done  by adding the observed number of events  in 
the corresponding cone which contains 90\% of the flux for a given
mass from  the Sun and the Earth, after scaling for their different
exposures. The same procedure is used to obtain the total  expected
background.  Using this, we obtain  the 90\% CL flux limit,
which is then scaled by the  cone collection efficiency.
We then evaluate  the direct detection rate at a given value of WIMP
mass and cross-section, using an exponential form factor~\cite{Freese88}, 
and using values for   the WIMP halo density and mean velocity from  
Ref.~\cite{CDMS}. Thus, for each value of WIMP mass, using the 
combined Super-K flux limit, the direct detection rate, and the maximum
expected ratio for the direct to indirect detection rate, Eqn.~\ref{ddrate}
is solved for $\sigma$ to get the 90 \% upper limit on the
WIMP-nucleon cross-section.  These Super-K upper limits on WIMP-nucleon 
scalar  cross-section are shown in Fig.\ref{figsksi} along with
CDMS~\cite{CDMS04}, EDELWEISS~\cite{Edelweiss}, ZEPLIN~\cite{ZEPLIN}   upper limits,  and DAMA~\cite{DAMA00} best
fit region. The dip  around a WIMP mass of 56 GeV
occurs because of  an enhancement in the WIMP capture rates in the
Earth caused by a  resonance due to WIMP mass matching  that of iron~\cite{Gould87}, whose abundance in the Earth is about 30\%~\cite{Anderson}.
There are two discontinuities in the Super-K  limits on WIMP-nucleon 
cross-section in Fig.~\ref{figsksi}. The first discontinuity occurs at around 80 GeV and 
the other occurs at around 174 GeV.  The reason for this is that these
limits have been calculated assuming a lower limit for $\xi(M)$, where
 $\xi$ is defined in Eqn.~\ref{xi}. For WIMPs more massive than the top quark
(with mass of around 174 GeV), the lower limit is given by WIMP
annihilation to  gauge bosons, and for WIMPs less massive than the   
top quark, the lower limit is given by WIMP annihilation to 
tau leptons. This continues until the WIMP mass is less  than the 
W boson mass at around 80 GeV, below which the lower limit comes
from WIMP annihilation to $b\bar{b}$ pairs. 
The Super-K limits rule out 
a significant portion of the WIMP parameter space favored by the
DAMA experiment with a very different technique.

\begin{figure}[hbt]
\center{\includegraphics[width=7.5cm]{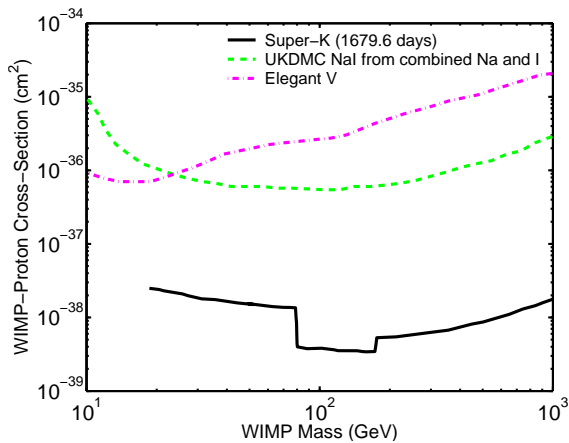}}
\caption{Super-K 90 \% CL exclusion region in WIMP parameter space
for a WIMP with spin-dependent coupling  along with corresponding 90 \%
CL exclusion limits from UKDMC (dashed)  and ELEGANT (dot-dashed).}
\label{figsksd} 
\end{figure}

To get limits on WIMP-proton spin-dependent cross-section, we carried
out the same exercise as above, this time using only the
limits from the Sun in the denominator of Eqn.~\ref{ddrate}, and using
the maximum inferred  ratio for direct-to-indirect detection 
rates with axial-vector couplings on the left hand side of Eqn.~\ref{ddrate}.
The Super-K limits on WIMP-proton cross-section are shown in
Fig.~\ref{figsksd} along with limits from other direct experiments like
UKDMC~\cite{UKDM} and ELEGANT-V~\cite{Elegant} experiments. The reason for the jumps around 80 GeV and 174 GeV 
is the same as that for the limits on WIMP-nucleon scalar
cross-section. The Super-K limits on WIMP-proton
cross-section are about 100 times more sensitive than those from 
direct detection experiments. 
  Also a limit on WIMP-proton spin-dependent cross-section using
earlier Super-K limits from the Sun~\cite{Okada}  has been done in
Ref.~\cite{Ullio}, which
 showed  that the annual modulation seen in DAMA  cannot be because 
of a WIMP with axial vector coupling to a proton,  since this is ruled out
by null searches for WIMP-induced annihilations in the Sun from Super-K
and other upward muon detectors.

It must be cautioned that these limits on WIMP-nucleon cross-section 
depend on the maximum expected ratio of direct-to-indirect detection
rates evaluated in Ref.~\cite{marc96}. These 
calculations  give a rough idea of expected sensitivities of upward muon 
detectors as compared to direct detection 
experiments. For WIMPs with significant
annihilation branching ratio to pure Higgs bosons or gluons,
the ratio could fall outside the indicated range~\cite{marc96}. 
Also, in  the comparison of direct and indirect detection rates,  oscillations
of neutrinos produced from WIMPs is not taken into 
account~\cite{Kowalski,Fornengo}.
Furthermore, the ratio does not include contribution  from the  proposed
bound solar system  population of WIMPs~\cite{Damour1,Damour2}. 
All these effects could change
the  limits from what we have evaluated. However, by using the maximum expected value
for the estimated ratio of direct to indirect detection rates in 
Ref.~\cite{marc96} we have obtained  conservative limits 
on WIMP-nucleon cross-section.

\section {Conclusions}
 We have looked for  indirect  signatures of dark matter using 1892
neutrino-induced upward through-going muon events by looking for
an  excess in the direction of the Sun, the Earth, and 
 the Galactic Center. We looked for an excess of
 upward muons over atmospheric neutrino  background  close to the centers
 of the above bodies. No statistically significant excess was seen.

  Flux limits were obtained for various cone angles around these potential
sources, and compared with previous estimates by other detectors.
These flux limits were calculated as a function of the WIMP mass. 

Then, using model-independent comparison of fluxes of direct to indirect 
detection rates obtained in Ref.~\cite{marc96}, we obtained limits on 
WIMP-nucleon cross-section for a WIMP with both scalar as well as axial-vector
couplings. \\

\section{Acknowledgments}
We gratefully acknowledge the cooperation of the
 Kamioka Mining and Smelting Company. The Super-Kamiokande experiment
 has been built and operated from funding by the Japanese Ministry of
 Education, Science, Sports and Culture, the  United States Department  of 
Energy, and the U.S. National Science Foundation, with support for 
individual researchers from Research Corporation's Cottrell College Science Award.

\end{document}